\documentstyle[amssymb,prb,aps,epsfig,multicol]{revtex}

\begin{document}
\title{Spin Fluctuations and the Magnetic Phase Diagram of ZrZn$_{2}$}
\author{I.I. Mazin and D.J. Singh}
\address{Code 6390, Naval Research Laboratory, Washington, DC 20375, U.S.A.}
\date{\today}

\maketitle

\begin{abstract}
The magnetic properties of the weak itinerant
ferromagnet ZrZn$_2$ are analyzed using Landau
theory based on a comparison of
density functional calculations and experimental data as a function
of field and pressure.
We find that the magnetic properties are strongly affected by
the nearby quantum critical point, even at zero pressure; LDA calculations
neglecting quantum critical spin fluctuations overestimate the
magnetization by a factor of approximately three.
Using renormalized Landau theory, we extract
 pressure dependence of the fluctuation amplitude. It appears 
that a simple scaling based on the fluctuation-dissipation theorem 
provides a good description of this pressure dependence.
\end{abstract}

\pacs{75.40.Cx,75.50.Cc,74.20.Rp}

\begin{multicols}{2}
The physics of metals near ferromagnetic quantum critical
points (QCPs) has attracted renewed interest following several
recent discoveries of materials with unusual still  poorly understood
transport and thermodynamic properties, as well
as unusual low-temperature states, particularly superconductivity
coexisting with ferromagnetism. \cite{saxena,aoki,pfleiderer}
From a generic, qualitative point of view, the phenomena
are understood as being connected with renormalization, scattering
and pairing due to strong fluctuations in the ferromagnetic
order parameter ({\em i.e.,} spin fluctuations) as the
critical point is approached.
Besides the mentioned
superconducting transition, in clean samples of many
weak itinerant
ferromagnetic metals  
the magnetic transition near the QCP crosses over from
the second order to the  weakly first order,
although whether this happens in ZrZn$_2$
is not yet established. The physics of this cross-over is not clear yet.
In any case,
quantitative, material specific understanding of these
phenomena is still lacking.

ZrZn$_{2}$ is a prototypical example of a weak itinerant (Stoner)
ferromagnet. Very small magnetic moments (0.12 to 0.23 $\mu _{B}$) have been
reported. These do not saturate even at fields up to 35 T, indicating
softness of the magnetic moment amplitude and suggesting existence of soft
longitudinal spin fluctuations. The Curie temperature, $T_C$ drops approximately
linearly with pressure, starting at $\approx 29$ K at $P=0$ and decreasing
to $\approx 4$ K at $P=16$ kbar,
\cite{uhlarz} which extrapolates to a
QCP at $P=18-20$ kbar. The discovery of superconductivity
in the ferromagnetic (FM) phase \cite{pfleiderer}
resulted in renewed interest
in this compound, including several theoretical studies
(Refs. \onlinecite{santi,singh2002,bruno,walker,powell} and others).
The relative structural simplicity of this compound and
the availability of high quality experimental data as functions
of $H$, $T$ and $P$ on clean samples suggest this material
as a test case for developing understanding of quantum critical
phenomena in ferromagnetic metals. Here we focus on the magnetic
properties, in particular the renormalization of LDA
results due to fluctuations.

Density functional theory is in principle an exact ground state
theory. It should, therefore, correctly describe the spin density
of magnetic systems. This is usually the case in actual
state of the art density functional calculations. 
However, common approximations to the exact density
functional theory, such
as the LDA, may miss important physics and indeed fail to describe
some materials.
A well know example is in
strongly Hubbard correlated systems, where the
LDA treats the correlations in an orbitally averaged mean field way and
underestimates the tendency towards magnetism.
Overestimates of magnetic tendencies, especially in the LDA,
are considerably less common, the exceptions being
materials near magnetic QCPs; here the error comes
from neglect of low energy quantum spin fluctuations.
Indeed, the LDA is
parameterized based on the uniform electron gas
at densities typical for atoms and
solids. However, the uniform electron gas at these densities is
stiff against magnetic degrees of freedom and
far from magnetic QCP's.
Thus, although the LDA is exact for the uniform electron gas,
and therefore does include all fluctuation effects in the
uniform electron gas, its description of magnetic ground states in
solids and molecules is mean field like.
This leads to problems
such as the incorrect description of singlet states in molecules
with magnetic ions as well as errors in solids
when spin fluctuation effects
beyond the mean field are important.
In solids near a QCP,
the result is an overestimate of the magnetic moments and tendency
toward magnetism ({\it i.e.} misplacement of the
position of the critical point) due to neglect
of the quantum critical fluctuations.
\cite{yamada,millis}
Examples include Sc$_3$In, \cite{aguayo} ZrZn$_2$,
\cite{singh2002}
and Sr$_3$Ru$_2$O$_7$ (Ref. \onlinecite{singh-327}).
Sr$_3$Ru$_2$O$_7$ displays a novel metamagnetic quantum critical point,
\cite{grigera}
while, as mentioned,
ZrZn$_2$ shows coexistence of ferromagnetism and superconductivity.
The effects of such quantum fluctuations can be described on a
phenomenalogical level using a Ginzburg-Landau theory,
in which the magnetic properties defined by the LDA fixed spin
moment curve are renormalized by averaging with an assumed
(usually Gaussian) function describing the beyond LDA critical
fluctuations. \cite{shimizu,larson} Although a quantitative theory
allowing extraction of this function from first principles calculations
has yet to be established, one can make an estimate based
on the LDA fixed spin moment curves as compared with experiment.

LDA calculations of the magnetic properties of ZrZn$_2$
(Refs. \onlinecite{santi,singh2002,bruno,huang})
are found to be sensitive to shape approximations,
possibly because of the very small energy scales involved.
In particular, atomic sphere approximations
result in smaller magnetizations than those found by more accurate
full potential methods. \cite{santi,singh2002,bruno}
In fact, it was found that full potential calculations produce a Stoner factor
of $\approx 1.16,$ as opposed
to 1.01 in the atomic sphere calculations, indicating a stronger
tendency to magnetism in the full-potential calculations. \cite{huang}

Our full potential LDA calculations for ZrZn$_2$ at its
experimental volume yield a magnetization of 0.72 $\mu _{B}$
per formula unit --
3 to 4 times larger than experiment,
reflecting an crucial role for the renormalization of the magnetization
due to beyond LDA fluctuations, presumably associated with the QCP.

As mentioned, one can
incorporate such fluctuations into LDA
calculations by renormalizing the
Landau expansion for the free energy
with Gaussian spin fluctuations of a given r.m.s. amplitude
(see Refs. \onlinecite{shimizu,larson} and references therein).
The latter can be obtained
empirically, or estimated from the parameters of the band structure
with an ansatz to separate spin fluctuations included
in the LDA from those neglected, as discussed in Ref. \onlinecite{larson}.
Here we report
renormalized Landau functional calculations where one parameter, the r.m.s.
amplitude of the beyond LDA fluctuations at $P=0$ is taken as an adjustable
parameter, determined by comparison with the experimental $P=0$ magnetization,
and use it to describe, without further
empirical parameters, the pressure and field dependence of the
magnetic properties of ZrZn$_{2}$.

The LDA calculations were done using the general potential linearized
augmented planewave (LAPW) method. Local orbital extensions were included
to accurately treat high lying core states and avoid linearization
errors. \cite{singhbook,singh-lo}
The Hedin-Lundqvist exchange correlation function was used
with von-Barth-Hedin spin scaling. \cite{hl,hl2}
The valence and Zr semicore $p$ states were treated in a scalar
relativistic approximation,
while the core states were treated fully relativistically.
LAPW sphere radii $R$=2.1 a$_0$ were employed with a dimensionless basis set
cutoff $R K_{max}$=9. Brillouin zone samplings were done using the
special {\bf k}-points method, with 182 points in the irreducible $1/48$
wedge of the zone. Convergence tests were done, showing that these
parameters were adequate. For example, fixed spin moment calculations
at the experimental lattice parameter were done using up to 1300
points in the wedge, with very slight changes of
less than 0.01 $\mu_B$ in the magnetization.
Calculations at the experimental lattice parameter were also
done with a different sphere radius, $R$=2.45 a$_0$, again with
negligible changes in magnetization.

In order to construct the Landau expansion,
we did fixed spin moment calculations, determining the total
energy as a function of magnetic moment an volume, using seven
lattice parameters from 13.0 $a_0$ to 13.9 $a_0$ plus the experimental
lattice parameter of 13.9358 $a_0$. \cite{str}
The variation of the energy with volume yields a bulk modulus, $B$=1.0 Mbar,
which we use to set the pressure scale, \cite{press}
since there is no experimental
value in the literature to our knowledge.
Using this value, the QCP at $P=18-20$ corresponds to a volume compression
of 1.7-1.9\%.

\begin{figure}[tbp]\centerline{\epsfig{file=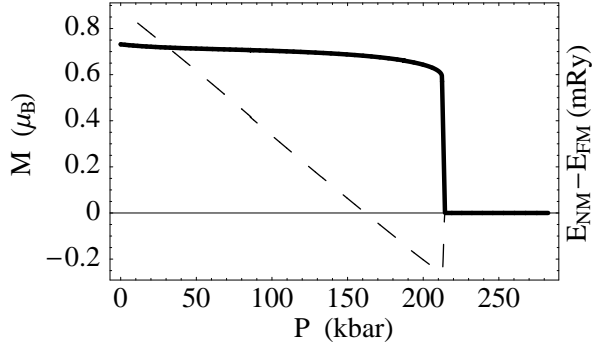,width=0.9\columnwidth}} 
\nopagebreak  
\caption{Unrenormalized LDA magnetic phase diagram: the solid line is the 
calculated magnetic moment for those pressures where a magnetic solution 
exists (left axis); the dashed line is the magnetic stabilization energy.
(right axis, same scale as left axis but units are mRy)
Note a 
metamagnetic behavior at $P\gtrsim 161$ kbar: there exists a magnetic solution,
although its energy is higher than that of the nonmagnetic state.
}
\label{mm}
\end{figure}

We now turn to the magnetic properties in the LDA.
As shown in Fig.\ref{mm}, the magnetization drops slowly from
0.72 $\mu _{B}$ at zero pressure ($V$=338 a$_0^3$) to 0.68 $\mu _{B}$
at $V$= 299 a$_0^3$, $P=161$ kbar (all volumes and magnetizations
are given per formula unit). At this pressure the ground state
becomes nonmagnetic and the moment
suddenly collapses to zero.
 The
ferromagnetic state remains metastable until
$V$=290 a$_0^3$, $P=$212 kbar.
Thus, the LDA predicts not a QCP, but a first order transition at
a pressure of $P\approx $ 161 kbar.
Leaving aside the question of the order of the transition,
the LDA strongly overestimates the magnetization, and has a much
higher transition pressure than experiment, implying an overestimate
of $T_C$ as well.
Additionally, the LDA yields
very weak $P$ dependence of the moment up to the transition
pressure, while experiment finds moments that decrease considerably with $P$
until at least $P\approx 16$ kbar. \cite{pfleid_unpb}

To proceed,
we use the fluctuation-renormalized Landau theory.\cite{note1}
A large literature exists on this subject, for instance, the review
of Ref. \onlinecite{shimizu}.
The basis of this theory is that the main
omission in LDA calculations is from long-range ferromagnetic spin
fluctuations, which are important near a QCP. One writes the
Landau expansion of the LDA total energy as 
\begin{equation}
E_{LDA}(M)=a_{0}+\sum_{n\geq 1}\frac{1}{2n}a_{2n}M^{2n},  \label{Hexp}
\end{equation}%
and then
introduces additional Gaussian zero-point fluctuations of an r.m.s. magnitude 
$\xi$ for each of the $d$ components of the magnetic moment (for a 3D
material $d=3$). After averaging over these, one
obtains a fluctuation-corrected functional. The general expression
\cite{shimizu,larson} reads
\begin{eqnarray}
E_{renormalized}(M) =a_{0}+&\sum_{n\geq 1}&\frac{1}{2n}\tilde{a}_{2n}M^{2n},  
 \label{renorm} \\
\tilde{a}_{2n} =\sum_{i\geq 0}C_{n+i-1}^{n-1}a_{2(n+i)}\xi ^{2i}&\Pi
_{k=n}^{n+i-1}&(1+\frac{2k}{d}).  \nonumber 
\end{eqnarray}

Two approaches are, in principle, possible at this point: one is to evaluate
$\xi$ using the fluctuation-dissipation theorem, 
\begin{equation}
\xi ^{2}=\frac{4\hbar }{\Omega }\int d^{3}q\int \frac{d\omega }{2\pi }\frac{1%
}{2}\mathop{\rm Im}\chi ({\bf q},\omega ),
\end{equation}%
however, this requires some knowledge of the susceptibility
$\chi({\bf q},\omega)$,
and a choice for the cutoff in the
integration. This choice gives the separation between
the fluctuations accounted for in LDA from
those missing. In the most pessimistic view it converts one
unknown parameter, $\xi$, into another, though it should
be said that the cutoff may be much less material and pressure
dependent than $\xi$ itself.
The other approach is to treat $\xi$ as an adjustable
parameter. Here we are interested in the magnetic phase diagram of
ZrZn$_{2}$ in a pressure range corresponding to that where magnetism is
observed experimentally, so it is possible to adjust $\xi$ to
reproduce the magnetic moment at ambient pressure and then use it for the
whole pressure range.
The fluctuation-dissipation theorem, though not used
directly, is used implicitly to construct an ansatz for
the $P$ dependence of $\xi ^{2}$: the
lowest-order expansion of the bare susceptibility
$\chi _{0}({\bf q},\omega)$,
\begin{equation}
\chi _{0}({\bf q},\omega )=N(E_{F})-aq^{2}+ib\omega /q  \label{8}
\end{equation}
gives rise, near a QCP, to the formula (see, {\em e.g.}, Refs.
\onlinecite{sol,kaul}), 
\[
\xi ^{2}=\frac{bv_{F}^{2}N(E_{F})^{2}}{8a^{2}\Omega }[Q^{4}\ln
(1+Q^{-4})+\ln (1+Q^{4})]
\]%
where $Q=q_{c}\sqrt{a/bv_{F}},$ $q_{c}$ is a cutoff in the momentum space, 
$\Omega$ is the Brillouin zone volume, and $v_{F}$ and $N(E_{F})$ are the
Fermi velocity and the density of states, respectively. The expression in
the square brackets depends on its argument logarithmically, so the
main volume dependence comes from the prefactor. Following
the arguments of Ref. \onlinecite{larson},
this prefactor scales with $V$ as
$V^{-1}$ in the effective mass approximation.
Thus, in the first approximation we write
$\xi ^{2}(V)=\xi ^{2}(V_{0})V_{0}/V$.

\begin{figure}[tbp]\centerline{\epsfig{file=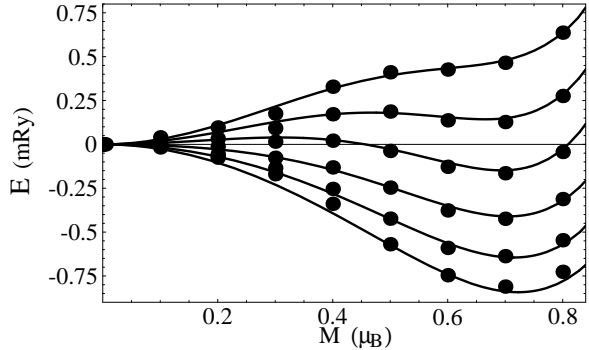,width=0.9\columnwidth}} 
\nopagebreak  
\caption{Fixed spin moment calculations for lattice parameters 
13.15, 13.30, 13.45, 13.60, 13.75, and 13.90 a$_0$. Solid lines are the 
6-th power fits according to Eq. \ref{Hexp}.}
\label{fits}
\end{figure}

\begin{figure}[tbp]\centerline{\epsfig{file=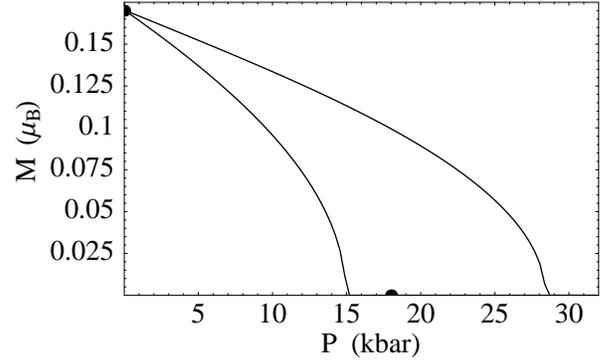,width=0.9\columnwidth}} 
\nopagebreak  
\caption{
Magnetization as a function of pressure, calculated from Eq.\ref{renorm},
using either a constant $\xi=0.5$ $\mu_B$ or with $\xi^2$ scaled as the inverse
cell volume (the right curve). Dots show the experimental magnetization
at zero pressure and the experimental critical pressure.}
\label{M(V)}
\end{figure}

In order to ensure stable fits, we have chosen the minimal power in
the expansion \ref{Hexp}, $n=6$. Fig.\ref{fits} shows the quality of the
fits, which is quite good. The value of $\xi (V_{0})$ that yields the
experimental value of the magnetic magnetization,
$M=0.17$ $\mu_B$, is then found to be
$\xi(V_{0})=0.5$ $\mu _{B}$.
Although the resulting dependence of $\xi$ on $V$ is
relatively weak, its effect on the phase diagram is large: In Fig. \ref{M(V)}
we show the (zero-temperature) equilibrium magnetization in zero field, as a
function of volume.
One can see that neglecting the volume dependence of
$\xi$ leads to a QCP at $P_{c}\approx 29$ kbar, while using the above
scaling, one gets nearly exact value $P_{c}\approx 15$ kbar. We should
recall, however, that this is the idealized phase diagram in zero field,
while actual measurements are performed in a small, but finite field. Near a
QCP even a small field can change magnetization drastically, as Fig.\ref%
{M(V)-H} illustrates. It is interesting to note that the metamagnetism
present in the bare
LDA calculations
disappears when the renormalization is included
account and as a result a QCP appears.
In reality, it may be that symmetry breakings other than
uniform ferromagnetism occur near the QCP and change the
transition to first order. It would be very interesting to 
experimentally investigate whether this in fact is the case, and
if so how close to the transition it occurs and what the relevant
order parameter is.

In summary, we report LDA calculations of the magnetic
energy of ZrZn$_{2}$ under pressure. Our results demonstrate that the LDA
substantially overestimates the tendency to magnetism in the whole
experimentally studied pressure range. This is an indication of
strong quantum spin fluctuations, associated with the QCP.
Using fluctuation-renormalized Landau theory, we
find that spin fluctuations with an  r.m.s. amplitude of 0.5 $\mu _{B}$
are needed at $P$=0 to obtain agreement with the experimental
magnetization. We further find that using a simple scaling based on the
fluctuation-dissipation theorem we are able to describe the phase
diagram up to the critical pressure with a good accuracy.

We are grateful for helpful discussions with G.G. Lonzarich,
C. Pfleiderer and S.S. Saxena. We thank C. Pfleiderer
for discussion of experimental data prior to publication.
Work at the Naval Research Laboratory is supported by the Office of
Naval Research. The DoD-AE code and
computing time from the DoD ASC and ARL centers
were used for some of the calculations.

\begin{figure}[tbp]\centerline{\epsfig{file=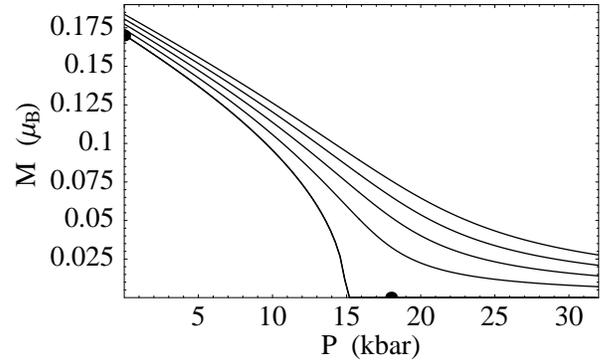,width=0.9\columnwidth}} 
\nopagebreak  
\caption{
Magnetization as a function of pressure, calculated 
with the scaled $\xi^2$, in an external fields of 0, 1, 2, 3, or 4 T. 
}
\label{M(V)-H}
\end{figure} 

\begin{figure}[tbp]\centerline{\epsfig{file=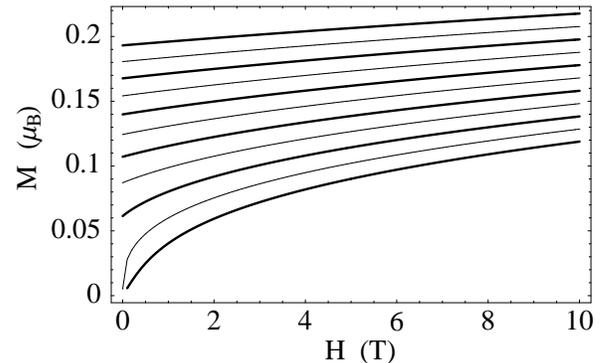,width=0.9\columnwidth}} 
\nopagebreak  
\caption{
Magnetization as a function of field;
the pressures are
0 to 20 kbar, spaced by 2 kbar, with alternating light and heavy lines.
}
\label{H}
\end{figure}

\end{multicols} 


\begin{references}
\bibitem{press}
Here we use the calculated $B$ to set the pressure scale with
the zero pressure point set to the experimental lattice parameter.
\end{references}
\end{document}